\documentstyle[twocolumn,aps]{revtex}

\begin{document}
\draft


%
\title{Adaptive coordinate, real-space electronic structure\\
calculations on parallel computers}
\author{Gil Zumbach, N. A. Modine and Efthimios Kaxiras}
\address{
Department of Physics,
Harvard University, Cambridge MA 02138 \\
(\today) \\
\begin{minipage}{5.5in}
\begin{abstract}
We present a method for electronic structure calculations that retains
all of the advantages of real space and addresses the inherent inefficiency
of a regular grid, which has equal precision everywhere.
The computations are carried out on a {\em regular} mesh in
{\em curvilinear space}, which allows
natural and efficient decomposition on parallel computers,
and effective use of iterative numerical methods.
A novel feature is the use of error analysis to
optimize the curvilinear grid for highly inhomogeneous electronic
distributions.  We report accurate all-electron calculations
for H$_2$, O, and O$_{2}$.
\end{abstract}
\end{minipage}
}
%
\maketitle
%
%


{\it Ab initio} electronic structure calculations are
computationally very challenging
because the singular Coulomb potential of the ions
results in highly localized core wavefunctions with
cusps at the ionic positions.
Even when pseudopotentials are used to eliminate
the core electrons, it is often desirable to treat valence
electrons with highly localized wavefunctions (e.g.,
1s, 2p, 3d, or 4f valence electrons), on the same footing as
delocalized ones.  Typical implementations use
a basis in a (one-particle) Hilbert space, the choice of which
requires a tradeoff between simplicity
and fast convergence of
physical quantities with the basis size.
The simplest basis consists of plane waves.
Its main drawback is uniform precision, leading
to slow convergence for
inhomogeneous systems like
atoms, molecules, clusters, or solid surfaces.
On the other hand, bases such as linearized augmented plane waves
(LAPW) or muffin tin orbitals (LMTO)
can be tailored to specific physical problems
and therefore have excellent convergence properties.
However, they lead to very complex equations.
A promising alternative is a real space approach.  All terms are
local except for the Laplacian, which has a very
short range.  The resulting sparse Hamiltonian allows effective
use of iterative algorithms, which vastly reduce both memory
and time requirements, and is a prerequisite to any $O(N)$ treatment of
electronic structure.

Harnessing the computational power of massively parallel architectures
imposes additional constraints on the choice of basis.
To achieve good load balance,
computational complexity and memory requirements must be evenly
divided among processors, a task made very difficult by
complex bases like LAPW and LMTO.
Another important consideration is the minimization
and localization of communication between processors.
Since Fourier transforms
(the underlying operations in a plane wave basis)
require communication between all processors,
even plane waves are not an efficient basis in this respect.
In contrast, a {\em regular} grid
in real space is a very natural choice for a massively
parallel computer architecture: assigning an equal section of the
grid to each processor provides good load balance,
minimizes interprocessor communication, and produces communication
patterns that are both local and conflict-free.

Chelikowsky and collaborators~\cite{Chelikowsky} have reported real
space electronic structure computations using a regular grid.
Recently, Briggs et al.~\cite{Briggs} have used multi-grid
acceleration to improve efficiency.
Yet, a regular grid in real space suffers from the same drawbacks
as a plane wave basis, i.e., it has the same resolution in
every region of space.  Attempts to circumvent this problem
have been pursued by Cho et al.~\cite{Cho93}
and by Wei et al.~\cite{Wei95}, using wavelets as a basis.
Another approach investigated by Tsuchida et al.~\cite{Tsuchida}
uses finite-elements with a non-uniform grid.
Finally, Bylaska et al.~\cite{Bylaska} have reported calculations
using multi-grid methods to enhance precision locally.
Irrespective of whether the formalism is based on wavelets,
finite-elements, or multi-grids, enhancing the resolution by
locally adding more basis elements ruins the natural mapping
onto a parallel architecture.

Progress toward overcoming the limitations of plane wave bases
has also been reported recently.
Gygi~\cite{Gygi} introduced the concept of adapted plane waves,
a distortion of Fourier space that
allows treatment of physical space with different degrees
of precision.
Following that development, Hamann~\cite{Hamann}
and Devenyi et al.~\cite{Devenyi94} reported
calculations using similar approaches.
The adaptive plane wave approach eliminates the major drawback
of the standard plane wave basis, but lacks the simplicity,
sparseness, and natural parallelization properties of real space algorithms.

In the present work, we combine the advantages of
real space calculations and those of adaptive coordinates in
a scheme for performing
electronic structure calculations in the context of
density functional theory and the local density approximation
(DFT/LDA)~\cite{HKS},
or the generalized gradient approximation~\cite{Perdew-Wang}.
Our scheme is
efficient and accurate for systems with very inhomogeneous
charge distributions and takes full advantage of
massively parallel computer architectures.
The central idea is to work on a regular grid,
but in a curvilinear space $\vec{\xi}$.
The change of coordinates $\vec{x}(\vec{\xi})$,
generates a {\em single} grid in real space $\vec{x}$,
which is finer where high precision is needed.
We refer to our method as the
Adaptive Coordinate, Real-space, Electronic Structure
(ACRES) algorithm.
It embodies the following advantageous features:

(1) It can achieve an essentially optimal distribution of grid points.
This is accomplished by a versatile choice
of the curvilinear coordinates, through which a grid of fixed size
is adapted to provide resolution commensurate with the physics.
The only cost of the adaptation
is the introduction of a nontrivial metric $g^{\alpha \beta}(\vec{\xi})$.

(2) The coordinate transform is chosen so as to minimize the
discretization error.
The idea is to determine an {\it a priori} good
set of curvilinear coordinates through the use of error analysis.
This differs from Gygi's original approach~\cite{Gygi}
in the adaptive plane wave scheme, where
the best possible change of coordinates is found
through an energy minimization.

(3) Since the communications pattern remains exactly the same as that
of a regular grid in real space, highly efficient parallelization
is trivially accomplished.

(4) The sparsity of the equations allows us to take advantage of
iterative algorithms.
This makes it possible to employ rather large grids,
and consequently to investigate complex systems.
The computational time scales as $N\times n_e$ with
$N$ the total number of points in the 3-dimensional
grid and $n_e$ the number of electrons in the system.

We will now describe the method in more detail.
The real space coordinates
$x^i(\xi^\alpha ; P^m)$
depend on
the curvilinear coordinates
$\xi^\alpha$ and on some set of parameters $P^m$ that
allow us to tune the change of coordinates
to a particular problem.
The Jacobian of the transformation is
\begin{equation}
J_\alpha^i(\xi; P) = \partial x^i/\partial \xi^\alpha
\end{equation}
with $|J| = \det J$ its determinant.
The trivial metric $g^{ij} = \delta_{ij}$ in real space
corresponds to
the metric
$g^{\alpha\beta} = {J^{-1}}_i^\alpha {J^{-1}}_i^\beta$
in curvilinear coordinates (summation over repeated indices implied).
The Laplacian operator in curvilinear space is
\begin{equation}
    \Delta = \frac{1}{|J|}\partial_\alpha\left
( |J| g^{\alpha\beta} \partial_\beta\right)
\label{Laplacian},
\end{equation}
and the integrals are transformed
according to $\int d^3x = \int d^3\xi\ |J|$.
The Coulomb potential is found by solving the Poisson equation
[discretized in curvilinear coordinates by means of the
Laplacian, Eq.~(\ref{Laplacian})] with the sum of the electronic
and nuclear charge as the source.

The equations are discretized in a box of linear size $\Lambda_i$,
using a finite difference scheme on a regular grid in curvilinear
space $\vec{\xi}$ with $N_{i}$ points in each direction~\cite{basis}.
Any boundary conditions, including the phase shifts required to
do multiple $k$-point calculations for solids, can easily be
implemented in this approach.  In the following, we use periodic
boundary conditions.

Implementation of the method presents certain challenges due to
the freedom in choosing discretization schemes.
The most important ones, and the manner in which we
resolved them, are discussed here briefly:

(a) Equations that are equivalent in the continuum limit
are not necessarily equivalent after discretization.
For example, there exist several
expressions for the Laplacian which are equivalent in
the continuum limit.  From physical and computational considerations,
it is desirable to have
a self-adjoint discretization of the Laplacian.
The expression given in Eq.~(\ref{Laplacian})
is self-adjoint after discretization if,
for a fixed pair of indices $(\alpha, \beta)$,
the finite difference operators used to
represent $\partial_\alpha$ and $\partial_\beta$ are identical.

(b) The order of the finite difference approximation
for the derivatives is very important.
The lowest order, two-point symmetric derivative is insufficient
and does not give good results.
Our experience indicates that we need to use a symmetric
discretization for the derivatives with at least four points (second order).

(c) The discrete representation of the nuclear charge is equally important.
For an atom with atomic number $Z$ at position $\vec{R}$,
the nuclear charge is $\rho(\vec{\xi}) = Z\,
\delta(\vec{\xi}; \vec{R})$ where
$\delta(\vec{\xi}; \vec{R})$
is a representation of a Dirac $\delta$ function at $\vec{R}$
on the regular grid in $\vec{\xi}$ space.
Beside the normalization condition on the $\delta$ function,
an important constraint on its representation
on a finite grid is that the first moment of the
distribution must correspond to the location of the $\delta$ function,
i.e.,
\begin{equation}
\int d\vec{\xi}\,|J|\;\delta(\vec{\xi};
\vec{R})\;\vec{x}(\vec{\xi}) = \vec{R}
\end{equation}
We found the most useful representation to be a Gaussian
\begin{equation}
\delta(\vec{\xi}; \vec{R}) \propto \exp{
\left( - |\vec{\xi} - \vec{\xi_0}|^{2}
\,/\;2\sigma^2\Delta\xi^2 \right)}
\end{equation}
with $\Delta\xi$ the regular grid spacing, $\sigma$ an adjustable parameter,
and $\vec{\xi}_{0}$ chosen to satisfy the constraint on the first
moment of the distribution.  This choice
reduces the translational invariance problems discussed in the next point.

(d) The presence of the grid breaks translational invariance.
We call the distance between an atomic center
and the nearest grid point the {\em offset}.
The energy depends on the offset,
and this effect can be quite large due to the Coulomb singularity.
The dependence is reduced by strong adaptation,
which makes the cell of the real space grid very small near the atomic
sites.
The Gaussian representation of the $\delta$ function
further reduces the dependence of the energy on the offset
because it results in a smoother transfer of charge as
the position of an atom changes.
The combined use of strong adaptation and a Gaussian $\delta$ function
eliminates the translation invariance problem.

(e) A final challenge is the actual choice
of curvilinear coordinates.
A necessary condition for the mapping
between $\vec{x}$ and $\vec{\xi}$
is that it must be
one to one, i.e., the grid in $x$ space must not be folded.
As the Laplacian involves the derivative of the metric,
and the metric is computed from the Jacobian,
the mapping must be at least $C^2$ on the torus
in order to ensure smoothness.
It is also desirable that the mapping be
spherically symmetric around an atom.
We use a two level coordinate transformation
$\vec{x}(\vec{\xi}; P)$ with a {\em global backdrop}
useful for simulating isolated structures,
and further {\em local adaptation} around each atom position.
The global backdrop is a simple independent
transformation along each axis $x^i = x^i(\xi^i)$ and
creates a flat central region with a high density of grid points
and a surrounding region with a
decreasing density of grid points.
The local adaptation creates a spherical deformation
of the grid around each atomic center $\vec{R}_\nu$,
with the amount of adaptation $A_\nu$ and the size of the adapted
region $\rho_\nu$ as independent variables.
As suggested by Gygi~\cite{Gygi}, for a given $\det{J(\vec{R}_\nu)}$
and $\rho_\nu$, the final results are not very sensitive
to the details of the formula for $\vec{x}(\vec{\xi})$.
The computations presented below were carried out
with the simple form for the local adaptation
\begin{equation}
  \vec{x}(\vec{\xi}; P) = \vec{\xi} -
\sum_\nu A_\nu\,(\vec{\xi} - \vec{R}_\nu )\,
        \exp\left(-\frac{|\vec{\xi} -
\vec{R}_\nu |^2}{2\,\sigma^2(A_\nu,\rho_\nu)}\right) \\
\end{equation}
with the function $\sigma(A,\rho)$ chosen such that $\rho$ gives
the real space width of the adapted region.

A question of central importance is how to choose the
different parameters of the grid so as
to generate a nearly optimal mesh for a given physical problem.
We resolve this issue by constructing an estimate
for the error in the integrals and then choosing the
parameters that minimize the error.  To illustrate this point,
consider a periodic, one-dimensional integral
$I(f) = \int d\xi\,f(\xi)$
computed numerically on a regular mesh
  \begin{equation}
    I_N(f) = \sum_i \Delta\xi\,f(\xi_i)
  \end{equation}
with $\Delta\xi = \Lambda/N$.
We evaluate the elementary error by comparing
the integrals computed with $N$ and $N/2$ points~\cite{euler}.
More precisely, with $N/2$ points,
the rectangular element of integration is
   $\delta I_{N/2} = 2 \Delta\xi\, f(\xi_i)$.
In comparison, the same element of integration computed with $N$ points is
   $\delta I_{N} = \Delta\xi\, [f(\xi_{i-1})/2 +
   f(\xi_{i}) + f(\xi_{i+1})/2]$.
An elementary estimate of the error is given by
\begin{equation}
\delta e(f)
  = \delta I_{N} - \delta I_{N/2}
  = \Delta\xi^3\, f''_{i}/2.
\end{equation}
If the constant 1/2 on the right hand side is replaced by 1/12
we obtain a rigorous upper bound due to Peano~\cite{Peano}.
The error in the numerical integral is then
estimated by the $L_2$ norm of $\delta e$
  \begin{equation}
   e(f) = \frac{1}{2}\left(\frac{\Lambda}{N}\right)^{5/2}
   \left( \sum_i \Delta\xi\, (f''_i)^2\right)^{1/2}.
  \end{equation}
The above idea is easily generalizable to
three-dimensional integrals
(whereas the rigorous Peano bound is difficult
to extend to higher dimensions).
The last step is to pick an integrand
$f$ so as to obtain an {\it a priori}
estimate of the optimal grid parameters by minimizing $e(f)$.
By experimenting on several atoms we have found that
$f = |J|\,\rho\,V_{K-S}$ provides an adequate indicator.
Due to large cancellations forced by the Kohn-Sham eigenvalue
equation, this term gives the leading factor for the error
in the total energy.


Using the approach described in this paper,
we have implemented DFT/LDA~\cite{Perdew-Zunger}
and DFT/GGA~\cite{Perdew-Wang}
electronic structure calculations
on the CM-5 massively parallel supercomputer.
Within this approach, all-electron
computations involving atoms in the first row
of the periodic table are feasible.
We have also implemented the pseudopotential approach,
using the norm-conserving nonlocal pseudopotentials
of Bachelet et al.~\cite{Bachelet82}, and the Kleinman-Bylander
procedure to render the nonlocal components
separable~\cite{Kleinman82}.

For the all-electron calculations, the adaptation of
the grid is determined by the requirement that
the density of grid points near the atomic cores is sufficient to
accurately represent the $1/r$ divergence of the Coulomb potential.
For example, Fig.~\ref{grid} shows
a grid used for the H$_{2}$ molecule calculation.
This clearly indicates the very large difference between the
spacing of grid points in the unoccupied vacuum region
and near the atomic nuclei.
Fig.~\ref{O2_wavefunctions} shows the occupied wave functions
of the O$_{2}$ molecule along a line through the centers of the
two atoms.
The enhancement of the grid resolution throughout the regions where
the electronic wave functions are large and the very
strong enhancement close to the nuclei
allow accurate representation of the smooth tails of the wave functions
as well as the cusps and nodes near the nuclei.

For a more quantitative comparison to other
theoretical results and to experiment,
Table~\ref{Summary_of_results} shows our
calculated results for H$_{2}$ and O.
It is clear from this comparison that our results are
in complete agreement with other theoretical work using similar
methods.  Therefore, the difference between
calculated values and
experimental measurements reflects fundamental limitations of the
underlying theory (DFT/LDA or DFT/GGA),
rather than limitations in the accuracy
of our method.

The authors are grateful to F.~Gygi for
sharing his insight on adaptive methods.
This work was supported
by the Office of Naval Research grant N00014-93-1-0190.

\begin{figure}
\caption{The $24\times 12\times 12$ [a.u.] grid used for H$_{2}$,
in a horizontal cross-section
through the atoms (every fourth line shown).  Notice the effect
of the global backdrop (crosslike pattern)
and the local adaptation around each atom.
}
\label{grid}
\end{figure}

\begin{figure}
\caption{Occupied wave functions of the O$_{2}$ molecule,
along a line through the centers of the atoms.
The $\pi$ bonding
and anti-bonding wave functions collapse onto
the horizontal axis (they have
nodes through the atomic centers).
The $1s$ bonding and anti-bonding states
were scaled by a factor of 1/3 so they could be displayed on
the same scale.
Points on the curves indicate values at actual
grid points used in the calculation.
}
\label{O2_wavefunctions}
\end{figure}

\begin{table}

\caption{Calculated
bond length $a_0$, vibrational frequency $\omega$, and minimum energy
$E_0$ for H$_2$ and atomic energy $E_{at}$ of O.
The zero-point vibrational energy has been subtracted from the experimental
total energy of H$_{2}$.}
\label{Summary_of_results}
\begin{tabular} {|l|c|c|c|}
 &
{\bf ACRES } &
{\bf Other DFT Theory } &
{\bf Experiment } \\
\hline
H$_2$ [LDA] & & & \\
$a_0$ (a.u.)  &
1.448  &
$1.446^{a}$ &
$1.401^{b}$ \\
$\omega$ (cm$^{-1}$)  &
4192 &
$4207^{a}$ &
$4401^{b}$ \\
$E_0$ (Ry) &
-2.276 &
$-2.27^{c}$ &
$-2.349^{b}$ \\
\hline
H$_2$ [GGA] & & & \\
$a_0$ (a.u.)  &
1.416  &
$1.413^{a}$  &
$1.401^{b}$ \\
$\omega$ (cm$^{-1}$)  &
4381 &
$4373^{a}$ &
$4401^{b}$ \\
$E_0$ (Ry) &
-2.340 &
$-2.34^{c}$ &
$-2.349^{b}$ \\
\hline
O [LDA] & & & \\
$E_{at}$ (Ry) &
-148.870 &
$-148.938^{d}$ &
$-150.027^{e}$ \\
\hline
O [GGA] & & & \\
$E_{at}$ (Ry) &
-149.912 &
$-149.994^{d}$ &
$-150.027^{e}$ \\
\end{tabular}
\vspace{0.2in}
${}^{a}$ Reference~\cite{Johnson92}, S-VWN and B-LYP \\
${}^{b}$ Reference~\cite{Huber79} \\
${}^{c}$ Reference~\cite{Perdew92}, LSD and PW GGA-II \\
${}^{d}$ Reference~\cite{Juan93}, LDA and PW91 \\
${}^{e}$ Reference~\cite{Moore71},
            Spin unpolarized ground state, $2p^{4}$ ${}^{1}$D\\

\end{table}


\begin{thebibliography}{10}

\bibitem{Chelikowsky}
J.~R. Chelikowsky, N. Troullier, and Y. Saad, Phys.\ Rev.\ Lett. {\bf 72},
  1240  (1994);
J.~R. Chelikowsky, N. Troullier, K. Wu, and Y. Saad, Phys.\ Rev.~B {\bf 50},
  11355  (1994).

\bibitem{Briggs}
E.~L. Briggs, D.~J. Sullivan, and J. Bernholc, preprint.

\bibitem{Cho93}
K. Cho, T.~A. Arias, J.~D. Joannopoulos, and P.~K. Lam, Phys.\ Rev.\ Lett. {\bf
  71},  1808  (1993).

\bibitem{Wei95}
S.~Q. Wei and M.~Y. Chou, Bull. Amer. Phys. Soc. {\bf 40},  43  (1995).

\bibitem{Tsuchida}
E. Tsuchida and M. Tsukada, Sol.\ St.\ Comm. {\bf 94},  5  (1995).

\bibitem{Bylaska}
E.~J. Bylaska {\it et~al.}, preprint.

\bibitem{Gygi}
F. Gygi, Europhys.\ Lett. {\bf 19},  617  (1992);
Phys.\ Rev.~B {\bf 48},  11692  (1993); {\bf 51},  11190  (1995);
Private communication.

\bibitem{Hamann}
D.~R. Hamann, Phys.\ Rev.~B {\bf 51},  7337  (1995); {\bf 51},  9508  (1995).

\bibitem{Devenyi94}
A. Devenyi, K. Cho, T.~A. Arias, and J.~D. Joannopoulos, Phys.\ Rev.~B
{\bf 49},  13373  (1994).

\bibitem{HKS}
P. Hohenberg and W. Kohn, Phys.\ Rev. {\bf 136},  B864  (1964);
W. Kohn and L. Sham, {\em ibid.} {\bf 140},  A1133  (1965).

\bibitem{Perdew-Wang}
J.~P. Perdew,  in {\em Electronic Structure of Solids '91}, edited by P.
  Ziesche and H. Eschrig (Akademie Verlag, Berlin, 1991).

\bibitem{basis}
Through our finite difference scheme, we have an approximation of the original
  equations, rather than a projection of the original problem onto a basis.
  Consequently, the variational aspect of a basis is lost, and our electronic
  eigenvalues are not necessarily upper bounds of the true electronic
  eigenvalues.

\bibitem{euler}
We caution the reader against a subtle but important pitfall in the error
  estimation of a numerical integral: the second Euler-Maclaurin summation
  formula is only an asymptotic formula.
  When applied to the case of a discretized integral of a $C^\infty$
  periodic function, it leads to an obvious contradiction!

\bibitem{Peano}
P.~J. Davis and P. Rabinowitz, {\em Methods of Numerical Integration}, 2nd  ed.
  (Academic Press, Orlando, FL, 1984).

\bibitem{Perdew-Zunger}
J.~P. Perdew and A. Zunger, Phys.\ Rev.~B {\bf 23},  5048  (1981).

\bibitem{Bachelet82}
G.~B. Bachelet, D.~R. Hamann, and M. Schl{\"{u}}ter, Phys.\ Rev.~B {\bf 26},
  4199  (1982).

\bibitem{Kleinman82}
L. Kleinman and D.~M. Bylander, Phys.\ Rev.\ Lett. {\bf 48},  1425  (1982).

\bibitem{Johnson92}
B.~G. Johnson, P.~M.~W. Gill, and J.~A. Pople, J.~Chem.\ Phys. {\bf 98},  5612
  (1992).

\bibitem{Huber79}
K.~P. Huber and G. Herzberg, {\em Molecular Spectra and Molecular Structure}
  (Van Nostrand Reinhold Company, New York, 1979), Vol.~IV.

\bibitem{Perdew92}
J.~P. Perdew {\it et~al.}, Phys.\ Rev.~B {\bf 46},  6671  (1992).

\bibitem{Juan93}
Y.-M. Juan and E. Kaxiras, Phys.\ Rev.~B {\bf 48},  14944  (1993);
Private communication.

\bibitem{Moore71}
C.~E. Moore, {\em Atomic Energy Levels} (U. S. Government Printing Office,
  Washington, 1971), Vol.~I.

\end{thebibliography}
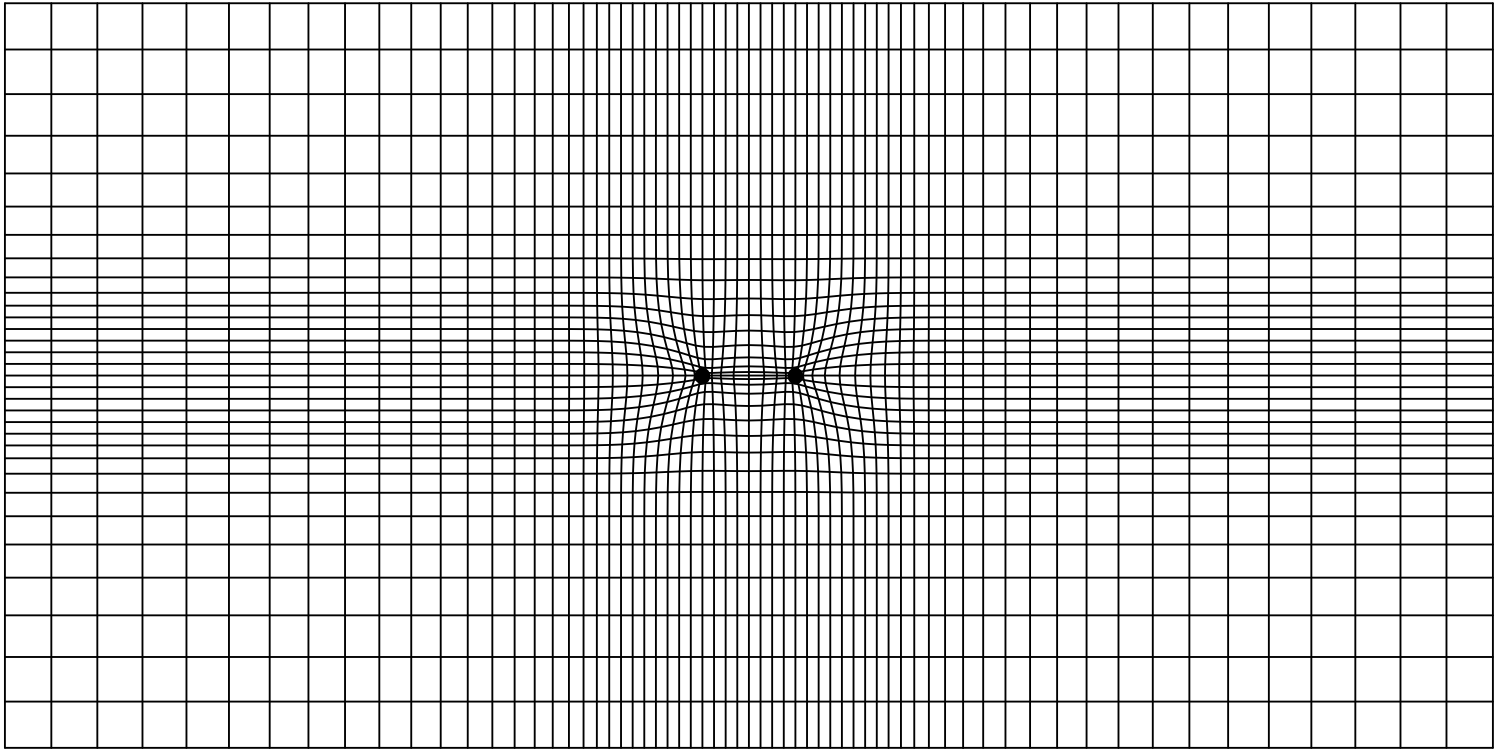
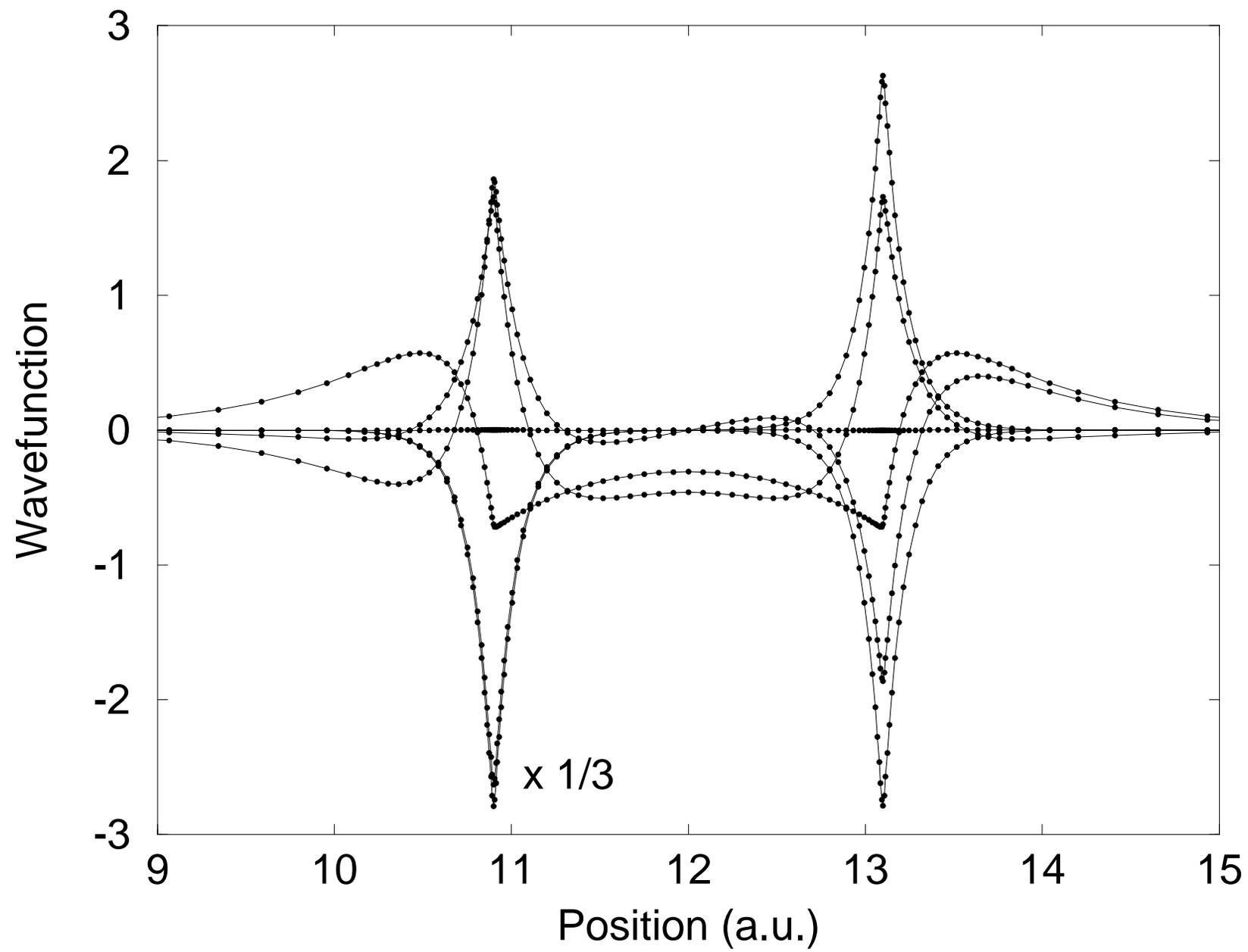
\end{document}